\documentclass[conference]{IEEEtran}

\IEEEoverridecommandlockouts

\usepackage{amsmath,amsthm,relsize}
\usepackage{amsfonts, amssymb, cuted}
\usepackage{cleveref}
\usepackage{mathtools}
\usepackage[]{algorithm}
\usepackage[noend]{algpseudocode}
\usepackage{cite}
\usepackage{xcolor}
\usepackage{soul}
\usepackage{graphicx}
\usepackage{tikz}
\usepackage{pgfplotstable}
\usepackage{pgfplots}
\usepackage{nicefrac}
\usepackage{enumitem}
\usepackage{support-caption}
\usepackage{subcaption}

\usepackage[font=footnotesize]{caption}
\usepackage{multirow}
\pgfplotsset{compat=1.18}
\usepackage{acro}
\usepackage{pifont}

\newtheorem{theorem}{Theorem}[]

\newtheorem{lemma}[theorem]{Lemma}

\theoremstyle{definition}

\newcommand{\CF}[0]{{\mathcal{F}}}

\newcommand{\CK}[0]{{\mathcal{K}}}
\newcommand{\CL}[0]{{\mathcal{L}}}

\newcommand{\CP}[0]{{\mathcal{P}}}

\newcommand{\CT}[0]{{\mathcal{T}}}

\newcommand{\Bu}[0]{{\mathbf{u}}}

\newcommand{\Bw}[0]{{\mathbf{w}}}
\newcommand{\Bx}[0]{{\mathbf{x}}}
\newcommand{\By}[0]{{\mathbf{y}}}
\newcommand{\Bz}[0]{{\mathbf{z}}}

\newcommand{\BH}[0]{{\mathbf{H}}}
\newcommand{\BI}[0]{{\mathbf{I}}}

\newcommand{\BW}[0]{{\mathbf{W}}}

\newcommand{\SfS}[0]{{\mathsf{S}}}

\usepackage{romannum}

\newcommand{\subparagraph}{}
\usepackage{titlesec}
\titlespacing\section{3pt}{6pt plus 4pt minus 2pt}{6pt plus 2pt minus 2pt}
\titlespacing\subsection{3pt}{4pt plus 4pt minus 2pt}{4pt plus 2pt minus 2pt}
\titlespacing\subsubsection{3pt}{3pt plus 4pt minus 2pt}{0pt plus 2pt minus 3pt}

\title{
Low-complexity Linear Multicast Beamforming for Cache-aided MIMO Communications
}

\begin{document}

\author{\IEEEauthorblockN{Mohammad NaseriTehrani, MohammadJavad Salehi and Antti T\"olli} \\
\IEEEauthorblockA{
    Centre for Wireless Communications, University of Oulu, 90570 Oulu, Finland \\
   \textrm{E-mail: \{firstname.lastname\}@oulu.fi}
    }
\thanks{This work is supported by the Academy of Finland under grants no. 346208 (6G Flagship) and 343586 (CAMAIDE).}
}

\maketitle

\begin{abstract}
A practical and scalable multicast beamformer design in multi-input multi-output~(MIMO) coded caching~(CC) systems is introduced in this paper. The proposed approach allows multicast transmission to multiple groups with partially overlapping user sets using receiver dimensions to distinguish between different group-specific streams. Additionally, it provides flexibility in accommodating various parameter configurations of the MIMO-CC setup and overcomes practical limitations, such as the requirement to use successive interference cancellation~(SIC) at the receiver, while achieving the same degrees-of-freedom~(DoF). 
To evaluate the proposed scheme, we define the symmetric rate as the sum rate of the partially overlapping
streams received per user, comprising a linear multistream multicast transmission vector and the linear minimum mean square error~(LMMSE) receiver.
The resulting non-convex symmetric rate maximization problem is solved using alternative optimization and successive convex approximation~(SCA). Moreover, a fast iterative Lagrangian-based algorithm is developed, significantly reducing the computational overhead compared to previous designs. The effectiveness of our proposed method is demonstrated by extensive simulations.

\end{abstract}

\begin{IEEEkeywords}
coded caching, multicasting, MIMO communications, finite-SNR performance
\end{IEEEkeywords}

\section{Introduction}
\label{section:intro}


\color{black} 
\par
Coded caching (CC) is a novel solution that benefits high-throughput, low-latency applications, such as extended reality~(XR), by leveraging the internal memory of networking devices to enhance system performance~\cite{salehi2022enhancing}. This is achieved through multicasting codewords to user groups of size $t+1$, 
where CC gain $t$ is proportional to the cumulative cache size in the network~\cite{maddah2014fundamental}. Additionally, CC approaches can be incorporated into multiple-input single-output (MISO) architectures, exploiting spatial multiplexing gain of $L$ at the transmitter to serve $t+L$ users in parallel and achieve a so-called degree-of-freedom (DoF) value of $t+L$~\cite{shariatpanahi2016multi,shariatpanahi2018physical}. Accordingly,~\cite{tolli2017multi} suggested using multi-group multicast optimized beamformers to enhance MISO-CC system performance at finite signal-to-noise ratio (SNR) while highlighting a trade-off between the number of overlapping multicast messages and spatial multiplexing gain for improved complexity and performance. 

\par Despite the significant research conducted on MISO-CC schemes, integrating CC in multi-input multi-output (MIMO) setups has not been explored as thoroughly. The DoF bounds for MIMO-CC systems with random cache placement in~\cite{cao2019treating}; however, achieving the bounds required complex interference alignment techniques. The study in~\cite{salehi2021MIMO} investigated low-complexity MIMO-CC schemes with a single transmitter and the spatial multiplexing gain of $G$ at each receiver, 
showing that the DoF of $Gt+L$ could be achieved when $\frac{L}{G}$ is an integer. Accordingly, to improve the finite-SNR performance of MIMO-CC systems, optimal unicast and relaxed multicast beamformer designs via alternative optimization and successive convex approximation (SCA) were developed in~\cite{salehi2023multicast}. The proposed designs, however, suffered from the computational complexity of successive interference cancellation (SIC) in receivers and limited flexibility in the design parameters. 
More recently, in~\cite{naseritehrani2023multicast}, we proposed a flexible MIMO-CC framework that optimized the DoF by a smart selection of the served users and improved upon the multicast beamformer design proposed in~\cite{salehi2023multicast} by formulating the symmetric rate problem with respect to (w.r.t) the transmit covariance matrices. 
While the resulting scheme enhanced the performance and relaxed the applicability criteria compared with~\cite{salehi2023multicast}, it may be  
too complex for practical implementation as it requires solving the rate optimization problem over the MAC region among multiple received partially overlapping multicast streams, and the rate constraints of the MAC region increase exponentially w.r.t the network size.
Additionally, both methods in~\cite{salehi2023multicast, naseritehrani2023multicast} rely on optimization-based tools, making implementation slow and non-scalable. 

\par This paper provides a practical and scalable extension to multicast beamformer designs in~\cite{salehi2023multicast,naseritehrani2023multicast} for MIMO-CC systems. We propose to use LMMSE receivers to separate the overlapping streams at the users, together with a linear multicast Tx beamformer design that builds upon the design in~\cite{mahmoodi2021low} but allows partially overlapping user groups. To approach the enhanced DoF introduced in~\cite{naseritehrani2023multicast}, we also advocate for a versatile transmission vector, that offers a more adaptable structure compared to the conventional CC scheme of~\cite{shariatpanahi2016multi}. 
To this end, we formulate the symmetric rate as the sum rate of the partially overlapping streams received per user.
As the resulting problem is non-convex, we use alternative optimization and SCA to find the symmetric rate. The result is a fast, iterative Lagrangian-based algorithm, paving the way to a practical and scalable Tx-Rx beamforming implementation that surpasses the limitations of prior designs and is highly adaptable to diverse scenarios without compromising the DoF.
Extensive numerical simulations are used to verify the performance and scalability of the proposed scheme and compare it with the existing state-of-the-art solutions.

Throughout the text, we use the following notations. For integer $J$, $[J] \equiv \{1,2,\cdot \cdot \cdot,J\}$. 
Boldface upper- and lower-case letters indicate matrices and vectors, respectively, and calligraphic letters denote sets. $|\CK|$ denotes the cardinality of the set $\CK$, $\CK \backslash \CT$ is the set of elements in $\CK$ that are not in $\CT$. Other notations are defined as they are used in the text.

\section{System Model}
\label{section:sys_model}
\subsection{Network Setup}
\label{section:network setup}
\begin{figure}[t]
        \centering
        \includegraphics[height = 4cm]{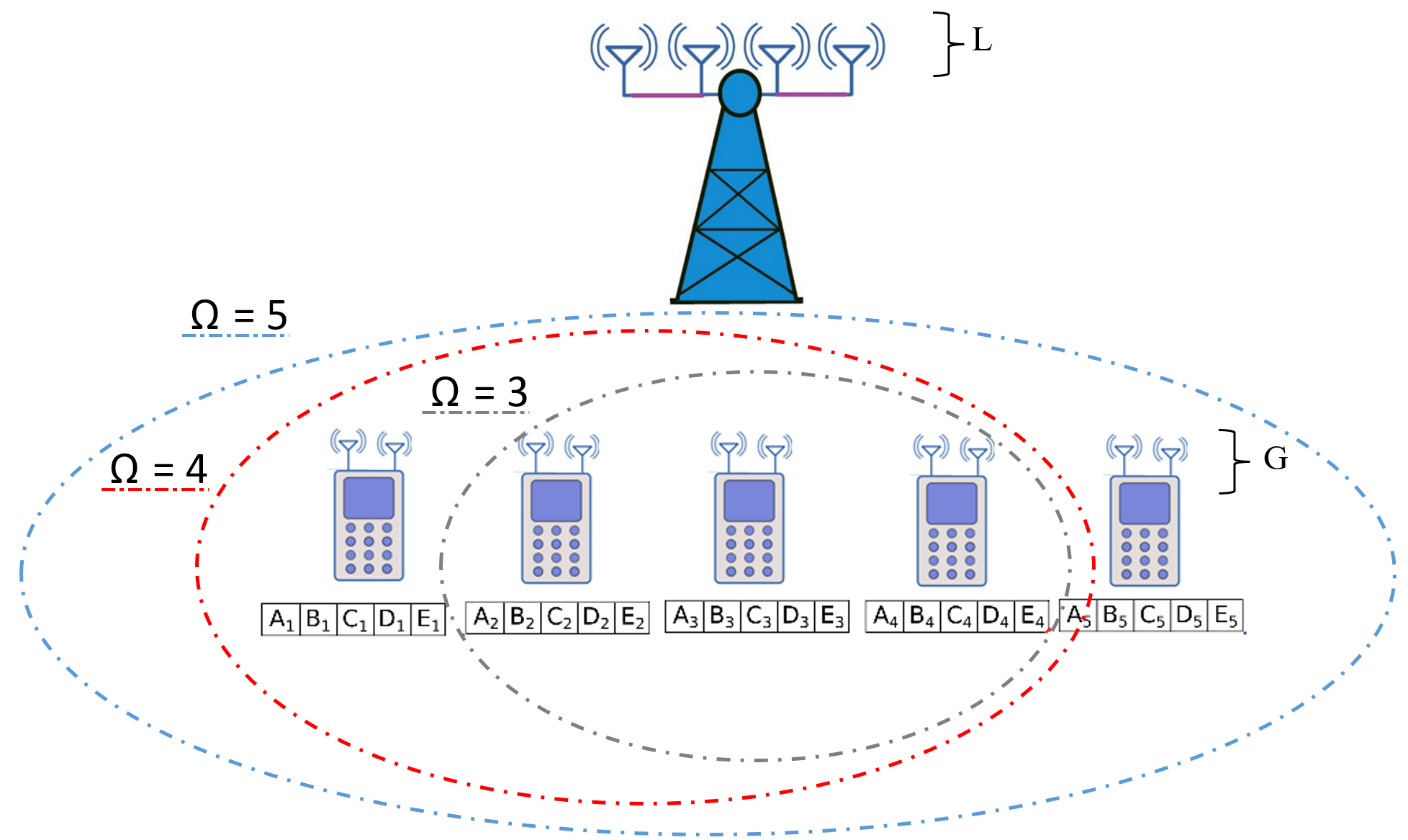} \label{fig:receiver_Prop}
    \caption{MIMO CC system model and the selection of users for different $\Omega$.}\label{fig:ISIT_sysm}
\end{figure}

We consider a MIMO setup where a single BS with $L$ transmit antennas serves $K$ cache-enabled users each with $G$ received antennas, as shown in Figure~\ref{fig:ISIT_sysm}.\footnote{
In fact, $L$ and $G$ represent the spatial multiplexing gain at the transmitter and receivers, respectively, which may be less than the actual number of antennas depending on the channel rank and the number of baseband RF chains. Nevertheless, the term `antenna count' is used for simplicity throughout the text.}
Every user has a cache memory of size $M$ data units and requests files from a library $\CF$ of $N$ unit-sized files.
The coded caching gain is defined as $t \equiv \frac{KM}{N}$, which represents how many copies of the file library could be stored in the cache memories of all users. 
The system operates through two main phases: placement and delivery. During the placement phase, the cache memories of users are loaded with data. Following a structure similar to~\cite{shariatpanahi2016multi}, we partition each file $W \in \CF$ into $\binom{K}{t}$ subfiles $W_{\CP}$, where $\CP \subseteq [K]$ represents any subset of users with $|\CP| = t$. Subsequently, in the cache memory of every user $k \in [K]$, we store $W_{\CP}$ for all $W \in \CF$ and all $\CP : k \in \CP$.

At the beginning of the delivery phase, each user $k$ reveals its requested file $W(k) \in \CF$ to the server. The server then constructs and transmits (e.g., in consecutive time slots) a set of transmission vectors $\Bx(i) \in \mathbb{C}^L$,
where $i \in [\binom{K}{\Omega}]$. Here, $\Bx(i)$ delivers segments of the requested data to every user in a subset $\CK(i)$ of users with $|\CK(i)| = \Omega$. The parameter $\Omega$ satisfies $t+1 \le \Omega \le t+L$ and represents the number of users served in each transmission. This choice is made to maximize the achievable DoF, as further clarified in Section~\ref{section:trans_vector_building}.
 

Upon transmission of $\Bx(i)$, user $k \in \CK(i)$ receives
\begin{equation}
\label{eq:RX_signal}
\By_k(i) = \BH_k \Bx(i) + \Bz_k(i) \; ,
\end{equation}
where $\BH_k \in \mathbb{C}^{G \times L}$ is the channel matrix between the server and user $k$, and $\Bz_k(i) \sim \mathcal{CN}(\mathbf{0},N_0 \mathbf{I})$ represents the noise. The entries of $\BH_k$ are assumed to be independent identically distributed (i.i.d) Gaussian variables with zero-mean and unit variance, and full channel state information (CSI) is considered available at the server. 

To calculate the symmetric rate, it is essential to determine the length (in data units) of each transmission vector. Following a similar rationale as in~\cite{shariatpanahi2018physical}, to ensure the delivery of new data in each transmission, we further partition every subfile $W_{\CP}$ into $\binom{K-t-1}{\Omega-t-1}$ equitably sized subpackets $W_{\CP}^q$. Additionally, as explained in Section~\ref{section:trans_vector_building}, the codeword construction process during the delivery phase involves XOR operations over a finite field without altering the data size. Consequently, the length of each transmission vector is equivalent to the subpacket length of $\frac{1}{\Theta}$, where $\Theta = \binom{K}{t}\binom{K-t-1}{\Omega-t-1}$ represents the \emph{subpacketization} level.
Now, utilizing $R_i$ to signify the multicast (max-min) transmission rate of $\Bx(i)$ that allows successful decoding at every user $k \in \CK(i)$, the transmission time of $\Bx(i)$ is denoted as $T_i = \frac{1}{\Theta R_i}$. The symmetric rate can then be defined as:
\begin{equation}
\label{eq:symmetric_rate}
    R_{sym} = \frac{K}{\sum_i T_i}= \frac{K}{\sum_i {\frac{1}{\Theta R_i}}} =  \frac{K {\binom{K}{t}\binom{K-t-1}{\Omega-t-1}}}{{\sum_i{\frac{1}{R_i}}}}  \; .
\end{equation} 
The goal is to design the delivery scheme to maximize $R_{sym}$. 

\subsection{Building the Transmission Vectors $\Bx(i)$}
\label{section:trans_vector_building}
The proposed cache placement strategy in Section~\ref{section:network setup}, which is a straightforward adaptation of~\cite{shariatpanahi2016multi}, enables us to create an XOR codeword for every subset $\CT$ of users with $|\CT| = t+1$, such that the data intended for every user is available in the cache memory of all the other $t$ users of $\CT$. As a tiny example, in a SISO setup with two users and the CC gain of $t=1$, if users~1 and~2 request files $A$ and $B$, respectively, we may simply broadcast $A_2 \oplus B_1$, where $\oplus$ denotes the XOR operation over a finite field. Then, by definition, user~1 has $B_1$ in its cache memory and can remove it from the received signal to decode $A_2$, and, similarly, user~2 can remove $A_2$ to decode $B_1$.


In our general MIMO-CC setup, we build a separate transmission vector $\Bx(i)$ for every subset $\CK(i)$ of users with $|\CK(i)| = \Omega$.  Without loss of generality, let us consider a single time interval
and remove the $i$ index for simplicity (the same process is repeated at every interval). In each slot, $ \Omega $ users are served with a transmission vector $\Bx$, where $\Omega$ is chosen to maximum achievable DoF~\cite{naseritehrani2023multicast},
such that:
    \begin{IEEEeqnarray}{lll}\label{eq:total_DoF}
        & {\mathrm{DoF}_{\max}} =  \max_{\beta , \Omega }~ \Omega \beta,   \nonumber\\
        &s.t.\:\:{\beta \le \mathrm{min}\Big({G},\frac{L \binom{\Omega-1}{t}}{1 + (\Omega - t-1)\binom{\Omega-1}{t}}\Big).} 
    \end{IEEEeqnarray}\\
Here, $\beta$ denotes the total number of parallel streams decoded by each user. The approach adopted in~\cite{naseritehrani2023multicast} to achieve this enhanced DoF 
requires solving the rate optimization problem over the
MAC region among multiple received partially overlapping
multicast streams, and the rate constraints of the MAC region
increases exponentially. To resolve this issue, we consider a linear stream-specific beamforming design to build $\Bx$, including multicast messages for every subset $\CT$ of $\CK$ with $|\CT| = t+1$. However, following the same intuition of SISO~\cite{maddah2014fundamental} and MISO~\cite{shariatpanahi2018physical} systems, the BS might be constrained to transmit at most $\binom{\Omega-1}{t}$ streams per user with the transmission vector $\Bx$ (this limit represents the number of partially overlapping streams over multicast groups in $\Bx$ that all include a specific user $k \in \CK$). This constraint limits the achievable DoF if $\beta > \binom{\Omega-1}{t}$. To address this issue,
we propose building the transmission vector $\Bx$ as follows:\\
\begin{IEEEeqnarray}{lll}\label{eq:Tx_vector}
    \Bx = \sum_{\CT \in \SfS^{\CK} } { {\BW}_{\CT} {\Bx}_{\CT }} ,
\end{IEEEeqnarray} 
where $\SfS^{\CK} = \{\CT \subseteq \CK, |\CT| =  t +1\}$ is the set of indices of multicast signals, $\BW_{\CT} = [\Bw_{\CT}^i]$, is formed by horizontal concatenation of vectors $\Bw_{\CT}^i$ over $i \in [q ]$, and $\Bx_{\CT} = [X_{\CT}^i]^T$ represents the column vector of $X_{\CT}^i$ constructed over $i \in [q ]$, $\forall \CT\in \SfS^{\CK}$. Here, ${X}_{\CT }$ is the multicast signal for user group $\CT \in \SfS^{\CK}$ from a 
complex Gaussian codebook. In~\eqref{eq:Tx_vector}, every ${X}_{\CT }$ is split into $q$ \emph{sub-streams} $X_{\CT}^i$, $i \in [q]$, and $\Bw_{\CT}^i$ is the corresponding Tx linear beamforming vector for $X_{\CT}^i$. The integer parameter $q$ is selected as the smallest value satisfying $q \binom{\Omega-1}{t} \geq \beta$.
Splitting every multicast signal ${X}_{\CT }$ into $q$ sub-streams allows us to increase the total number of parallel data streams to $q \binom{\Omega-1}{t}$, approaching the DoF bound of~\cite{naseritehrani2023multicast} if $\beta > \binom{\Omega-1}{t}$ (The bound is attained when $\beta$ is divisible by $\binom{\Omega-1}{t}$).
After transmitting $\Bx$, user $k \in \CK$ receives
\begin{equation}
\label{eq:RX_signal_lin}
\By_k = \BH_k  \sum_{\CT \in \SfS_k^{\CK} }  { {\BW}_{\CT} {\Bx}_{\CT }}
 + \BH_k \sum_{\CT \in \bar{\SfS}_k^{\CK} } { {\BW}_{\CT} {\Bx}_{\CT }}
 + \Bz_k \;.
\end{equation} 
where the first and the second summations represent the intended and interference terms for user $k$, respectively. Here, $\SfS_k^{\CK} = \{\CT \in \SfS^{\CK} \mid k \in \CT \}$ denotes the set of indices of multicast signals including data for user $k$.

\section{Multi-Group Multicast Transmission Design}
\label{section:Multi-Group Multicast Communication in Coded caching}
We examine the worst-case delivery rate among all users in $\CK$ to meet their requests for files from the library. 
We introduce a linear extension of the flexible multigroup multicasting transmission strategy for MIMO-CC  systems in~\cite{naseritehrani2023multicast}. Our objective is to optimize the sum rate across all the $q$ sub-streams of each stream. Given that $\binom{\Omega-1}{t}$ streams are requested each user,
we can formulate the problem of maximizing symmetric rates for the MIMO multigroup multicasting model in~\eqref{eq:RX_signal_lin} as follows: 
\begin{IEEEeqnarray}{lll}
    \label{eq:optmization_problem}
    \max_{\Bw_{\CT}, \Bu_{k,\CT}^i} ~&
 \min_{ k\in {\CK}}  \sum_{i \in [q ]} \min_{ {\CT}\in \SfS_k^{\CK} } \log(1 +  \gamma_{k,\CT}^i)   \nonumber\\ 
  s.t.~~&\sum_{\CT \in \SfS^{\CK}, i \in [q ] } \|\Bw_{\CT}^i\|^2  
	  \leq P_T ,
   \end{IEEEeqnarray}
where $\gamma_{k,\CT}^i$ is the SINR term at user $k \in \CT$, for decoding the $i$-th substream of the stream corresponding to $\CT$. To calculate $\gamma_{k,\CT}^i$, we first note that since $\beta \le G$, we can decode all the partially overlapping sub-streams without any need to SIC, using a simple LMMSE receiver. This is done by multiplying $y_k$ in~\eqref{eq:RX_signal_lin} by receive beamforming vectors 
\begin{IEEEeqnarray}{lll}
\label{eq:receive_beamformer}
\Bu_{k,\CT}^i = 
\Big(\BH_k \BW\BW^H \BH_k^H + N_0 \BI\Big)^{-1} \BH_k \Bw_{\CT}^i,  
\nonumber \\ \qquad \qquad \qquad \qquad \qquad \forall \CT \in \SfS_k^{\CK},  i\in [q], k\in \CK
\end{IEEEeqnarray}
where $\BW = [\Bw_{\CT}^i]$, $\CT\in \SfS^{\CK}, i \in [q ]$ is the concatenation of all beamforming vectors in $\Bx$, and $P_T$ is the transmission power. Now, we can calculate $\gamma_{k,\CT}^i$ as
\begin{IEEEeqnarray}{lll}
\label{eq:snr}
\hspace{-6pt}\gamma_{k,\CT}^i \!\!= \!\!\frac{  |{\Bu_{k,\CT}^i}^H \BH_k \Bw_{\CT}^i|^2}{ \sum_{\bar{\CT} \in \SfS_{\mathrm{}}^{\CK}, j\in [q] \backslash (\CT,i) }   |   {\Bu_{k,\CT}^i}^H \BH_k \Bw_{\bar{\CT}}^j|^2+ N_0 \|\Bu_{k,\CT}^i\|^2}.
\end{IEEEeqnarray}

One efficient way to solve the optimization problem~\eqref{eq:optmization_problem} is to use alternative optimization by solving for $\{\Bu_{k,\CT}^i\}$ or $\{\Bw_{\CT}^i\}$ assuming the other set is fixed, and updating the variables and solving for the other set iteratively until convergence. While for fixed $\{\Bw_{\CT}^i\}$, $\{\Bu_{k,{\CT}}^i\}$ is given simply by~\eqref{eq:receive_beamformer}, solving the other way around requires relaxing the non-convex problem, e.g., using SCA. In this paper,  
we propose a fast, iterative SCA solution that is built upon the design in~\cite{mahmoodi2021low} but allows partially overlapping streams. First, we rewrite the problem in the epigraph form:
\begin{IEEEeqnarray}{lll}\label{eq:epigraph}
 &\max_{\Bw_{\CT}^i, r_c , t_{k,\CT}^i, r_{k}^i}~ r_c\nonumber\\
    &s.t.\:\: r_c \leq \sum_{i \in [q]} r_{k}^i\:\:, r_{k}^i \leq  t_{k,\CT}^i,
    \:\nonumber\\  &\quad\epsilon_{k, \CT}^i \leq  \bar{\alpha}_{k, \CT}^i t_{k,\CT}^i+ \bar{\psi}_{k, \CT}^i,\quad \forall k\in \CK, \CT \in \SfS_k^{\CK}, i \in [q],  \nonumber\\
    &\quad \sum_{{\CT}\in \SfS^{\CK}, i \in [q] }  \|\Bw_{\CT}^i\|^2 \leq P_T,  
  \end{IEEEeqnarray}
where $\epsilon_{k, \CT}^i = (1+\gamma_{k,\CT}^i)^{-1}$ is the MSE term, and 
\begin{IEEEeqnarray}{lll}
\bar{\alpha}_{k, \CT}^i = \frac{f'(\bar{t}_{k, \CT}^i)}{f(\bar{t}_{k, \CT}^i)^2}, \qquad \bar{\zeta}_{k, \CT}^i = \frac{f(\bar{t}_{k, \CT}^i) + \bar{t}_{k, \CT}^i f'(\bar{t}_{k, \CT}^i)}{f(\bar{t}_{k, \CT}^i)^2}, \nonumber
\end{IEEEeqnarray}
for $f(t_{k,\CT}^i)=2^{t_{k,\CT}^i}$ (more details are provided in~\cite{kaleva2016decentralized}). As a result, the problem is solved iteratively via SCA and AO, as is also shown by the pseudo-code of Algorithm~\ref{F1}. 
\begin{algorithm}[t] 
	\caption{CVX-based solution for \eqref{eq:optmization_problem} }
		\small\begin{enumerate}
		\item \textbf{Initialize} 
		$t_{k,\CT}^{i}, \Bw_{\CT}^i, r_c , k \in \CK,  \CT \in {\SfS}_k^{\CK}, i\in [q] $.
        set $\Bw_\CT^i$ random vectors to satisfy power budget
		\item \textbf{Repeat} till convergence is met and ${i}_1 < I_{out}$, set ${i}_1 = {i}_1 + 1$, $i_1  = 0$; Compute ${\Bu_{k,\CT}^{i}}^{(0)}$ from~\eqref{eq:receive_beamformer} 
            \item \textbf{Repeat} till convergence in SCA and $i_2 < I_{in}$, set ${i}_2 = {i}_2 + 1$,
        \item Compute ${\epsilon_{k,\CT}^{i}}$, using $\Bw_\CT^i$, ${\Bu_{k,\CT}^{i}}^{(i_1-1)}$. 
        \item Update $\bar{t}_{k,\CT}^{i} = f(({\epsilon_{k,\CT}^{i}})^{-1})$, and $\bar{\alpha}_{k,\CT}^{i}$, $\bar{\zeta}_{k,\CT}^{i}$.
        \item  Solve \eqref{eq:epigraph} with CVX and find $\Bw_\CT^i$, $r_c$, $t_{k,\CT}^{i}$;\:\: 7)  \textbf{end}, 8) \textbf{end} 
	\end{enumerate}\label{F1}
\end{algorithm}

\par Now, for an even faster and more scalable implementation, we consider an alternative iterative approach implemented via the Lagrangian function. To do so, we can write the Lagrangian function of~\eqref{eq:epigraph} as
\begin{IEEEeqnarray}{lll}
     \hspace{-6pt} \CL(\lambda_{k, \CT}^i , t_{k, \CT}^i, \zeta_{k}, {v}_{k, \CT}^i, \mu, \{\Bw_{\CT}^i\}, r_c )  =  \nonumber \\     
     -r_c +  \mu \Big( \sum_{{\CT}\in \SfS^{\CK}, i \in [q]}  \|\Bw_{\CT}\|^2 - P_T \Big)  
   +   \sum_{{k}\in {\CK}  }  {\zeta}_{k} \Big(r_c - \sum_{{i}\in [q] } r_{k}^i\Big)  \nonumber\\
    +    \sum_{k\in \CK, \CT \in \SfS_k^{\CK}, i \in [q]}   {v}_{k, \CT}^i \Big(r_{k}^i - t_{k, \CT}^i \Big) \nonumber \\
    +  \sum_{k\in \CK, \CT \in \SfS_k^{\CK}, i \in [q] } {\lambda}_{k, \CT}^i  \Big(\epsilon_{k, \CT}^i - \bar{\alpha}_{k, \CT}^i t_{k, \CT}^i - \bar{\psi}_{k, \CT}^i\Big),\label{eq:P_optimum_Lagrang}
 \end{IEEEeqnarray} 
where the dual variables $\lambda_{k, \CT}^i$, ${v}_{k, \CT}^i$, and $\mu$ are associated with the MSE, common rate, and power, respectively.
\begin{lemma}
\label{lem:kkt}
For fixed receive beamformers $\{\Bu_{k,\CT}^i\}$, the optimal points for primal and dual variables in~\eqref{eq:P_optimum_Lagrang} fulfill:
\small\begin{align}
&{\Bw_{\CT}}^{i*}(\lambda_{k, \CT}^{i*} , \mu^*, \{\Bu_{k,\CT}^i\}) = \nonumber \\
&\quad \Big( \!\!\!\!\!\!\!\!\!\!\!\!\!\! \sum_{k\in \CK, \CT \in \SfS_k^{\CK},i\in  [q] }  \!\!\! \! \!\!\! \!\BH_k^H \Bu_{k,\CT}^i{\Bu_{k,\CT}^i}^H \BH_k + \mu \BI \Big)^{-1}   \sum_{ k \in {\CK} }  \lambda_{k, \CT}^{i*} \BH_k^H \Bu_{k, \CT}^i, \label{eq:primal_wg}\\
&r_{k}^{i*}({v}_{k,\CT}^{i*}, \epsilon_{k,\CT}^{i*}) =   \frac{\sum_{{\CT} \in \SfS_k^{\CK} }  {v}_{k,\CT}^{i*} \log((\epsilon_{k,\CT}^{i*})^{-1})}{\sum_{{\CT} \in \SfS_k^{\CK} }  {v}_{k,\CT}^{i*}},\label{eq:primal_rli}\\
&r_{c}^*({v}_{k,\CT}^{i*} , \epsilon_{k,\CT}^{i*}) = \! \! \! \sum_{{k}\in {\CK}, {i}\in [q] }  \! \!  \Bigg[\frac{\sum_{{\CT} \in \SfS_k^{\CK} }  {\zeta}_{k}^{*}{v}_{k,\CT}^{i*} \log((\epsilon_{k,\CT}^{i*})^{-1})}{\sum_{{\CT} \in \SfS_k^{\CK} }  {v}_{k,\CT}^{i*}}\Bigg], \label{eq:primal_rc}\\
&\frac{1}{q}\sum_{\CT \in \SfS_k^{\CK}, i \in [q] } {v}_{k,\CT}^{i*} = 1, \label{eq:dual_vkl_sum}\\
&\lambda_{k,\CT}^{i*} =  \frac{{v}_{k,\CT}^{i*}}{\epsilon_{k,\CT}^{i*} \log(2)}, \label{eq:dual_lambda}\\
& \mu^*(\lambda^{*} , \Bu_{k,\CT}^{i} )  =  \frac{1}{P_T}\sum_{k\in \CK, \CT \in \SfS_k^{\CK}, i \in [q] }  \lambda_{k,\CT}^{i*}  {\Bu_{k,\CT}^{i} }^H\Bu_{k,\CT}^{i}, \label{eq:dual_mu} \\
& {v_{k,\CT}^i}\!\!^{(i_2)} \!\! =\!\!   \Big({v_{k,\CT}^i}\!\!^{(i_2-1)}\!\!+ \eta \nabla_{v_{k,\CT}^i}\CL(.) \Big)^{+}\!\!, \label{eq:subgradient_vkl}\\
&\qquad \qquad \qquad \qquad \qquad \qquad \qquad \qquad \forall k\in \CK,\CT \in \SfS_k^{\CK},i \in [q]. \nonumber
\end{align}
\normalsize
\end{lemma}
The proof of Lemma~\ref{lem:kkt} is established from stationary KKT conditions and complementary slackness w.r.t primal and dual variables~\cite{mahmoodi2021low}. It is not feasible to propose a closed-form solution for the interdependent dual variables $v_{k, \CT}^{i} $; however, the sub-gradient method can still be applied to update $v_{k,\CT}^{i} $ and address this issue~\cite{kaleva2016decentralized}:
\begin{IEEEeqnarray}{lll}
\label{eq:subgradient_vkl_1}
{v_{k,\CT}^i}^{(i_2)} =   \Big({v_{k,\CT}^i}^{(i_2-1)}+ \eta \nabla_{v_{k,\CT}^i}\CL(.) \Big)^{+},
\end{IEEEeqnarray}
where $\nabla_{v_{k,\CT}^i}\CL(.)  = r_c + \log(\epsilon_{k,\CT}^i)$ and $[x]^{+} =\max(x,0)$. 
The pseudo-code for the proposed solution is provided in Algorithm~2.



\section{Numerical Examples}
\label{section:NumRes}

Numerical results are generated for various combinations of network and design parameters $L$ $G$, $t$, and $\Omega$ and for different transmission strategies. Channel matrices are modeled as i.i.d complex Gaussian, and the SNR is defined as $\frac{P_T}{N_0}$, where $P_T$ is the power budget and $N_0$ denotes the fixed noise variance. 
For a comprehensive analysis, we consider various baseline schemes, including the covariance-based design of~\cite{naseritehrani2023multicast} (upper bound) and zero-force (ZF) beamforming (lower bound).  

In Figure~\ref{fig:plot_1}, assuming $L =  \{3, 4\}$, $G  =  2$, $\Omega =  3$, and $t  =1$, 
we have compared the covariance-based design and ZF beamforming with different combinations of alternating optimization solutions, including CVX or the proposed KKT solution for transmit beamformers and MMSE-SIC or LMMSE solution for receive beamformers. 
As can be seen,  despite being less complex, the proposed LMMSE receivers perform well compared with both MMSE-SIC and the upper bound. It is also evident that the performance gap gets smaller if the system is not fully loaded, i.e., as $L$ is increased from $3$ to $4$ without changing $\Omega$. %
Therefore, by adopting the proposed method, one can effectively approach the upper bound performance, closely tracking it, all while maintaining the same DoF. 
Furthermore, the results confirm that the proposed iterative algorithm (KKT) provides the same solution as CVX while incurring notably reduced computational overhead.
\par
\begin{figure}[ht]
    \centering
    \resizebox{0.84\columnwidth}{!}{%

    \begin{tikzpicture}

    \begin{axis}
    [
    axis lines = center,
    xlabel near ticks,
    xlabel = \smaller {SNR [dB]},
    ylabel = \smaller {Symmetric Rate [bits/s]},
    ylabel near ticks,
    ymin = 0,
    xmax = 30,
    legend pos = north west,
    ticklabel style={font=\smaller},
    grid=both,
    major grid style={line width=.2pt,draw=gray!30},
    ]
    
    
    
    

    

        \addplot
    [dashed, mark = diamond , red!90]
    table[y=RsymLG63K420LMMSEKKT,x=SNR]{Figs/data.tex};
    \addlegendentry{\tiny KKT-LMMSE-$L= 6, G=3$ }

    \addplot
    [dashed, mark = o , red!90]
    table[y=RsymLG64K420LMMSEKKT,x=SNR]{Figs/data.tex};
    \addlegendentry{\tiny KKT-LMMSE-$L= 6, G=4$}

    \addplot
    [dashed, mark = square , blue]
    table[y=RsymLG73K420LMMSEKKT,x=SNR]{Figs/data.tex};
    \addlegendentry{\tiny KKT-LMMSE-$L= 7, G=3$}
    
    \addplot
    [dashed, mark = star , blue]
    table[y=RsymLG74K420LMMSEKKT,x=SNR]{Figs/data.tex};
    \addlegendentry{\tiny KKT-LMMSE-$L= 7, G=4$ }
    

 \addplot
    [dash dot, mark = star , brown]   table[y=RsymproposedLMMSEL6G4t0omg4,x=SNR]{Figs/data.tex};
    \addlegendentry{\tiny KKT-LMMSE-$L= 6, G=4$, t = 0 }


    
    
    \end{axis}

    \end{tikzpicture}
    }

    \caption{Proposed iterative solution for large DoF setup; $t=2, \Omega = 4, K=20$}
    \label{fig:plot_2}
\end{figure}
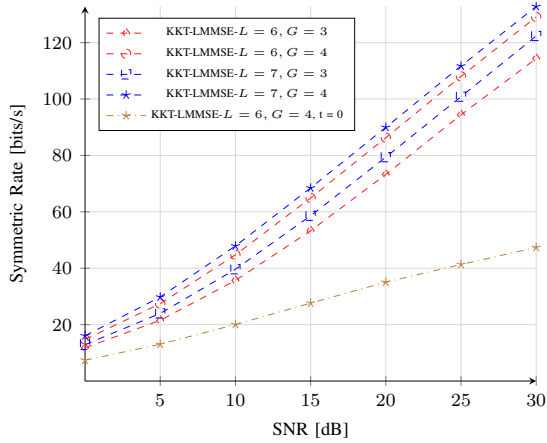

 In Figure~\ref{fig:plot_2}, we examine how the proposed solution could be applied to network setups with a possibly large DoF value.
For this simulation, we use LMMSE receivers and employ the proposed iterative design for transmit beamformers. For network parameters, we assume $L = \{6, 7\}$, $G = \{3,4\}$, $t = 2$, and $\Omega = 4$.
As can be seen, with the proposed solution, we can calculate the symmetric rate even when the achieved DoF gets as large as 12, which is much larger than what was possible with the methods presented in~\cite{salehi2023multicast,naseritehrani2023multicast}. This is a direct result of the low computational overhead of the proposed algorithms, which removes the need for CVX and allows high-performance beamformer designs when the number of parallel streams (i.e., DoF) is large. The results also show the performance gain as $L$ and $G$ increase from their minimum values of $6$ and $3$, respectively. Moreover, one can recognize there is a significant enhancement in both symmetric rate and DoF in comparison to conventional MIMO counterparts without coded caching, i.e., when $t = 0$.

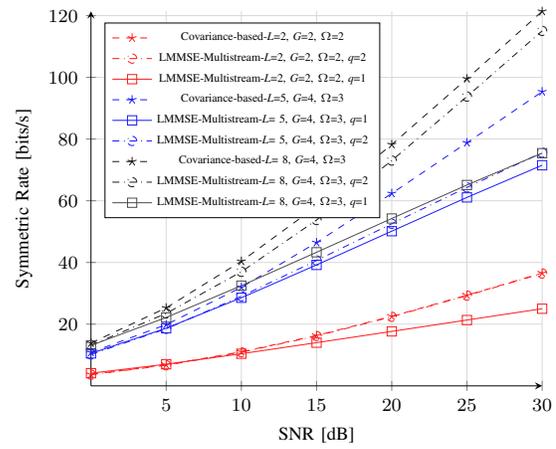
\begin{figure}[ht]
    \centering
    \resizebox{0.84\columnwidth}{!}{%

    \begin{tikzpicture}

    \begin{axis}
    [
    axis lines = center,
    xlabel near ticks,
    xlabel = \smaller {SNR [dB]},
    ylabel = \smaller {Symmetric Rate [bits/s]},
    ylabel near ticks,
    ymin = 0,
    xmax = 30,
    legend pos = north west,
    ticklabel style={font=\smaller},
    grid=both,
    major grid style={line width=.2pt,draw=gray!30},
    ]
    
    
    
    

    



     \addplot
    [dashed, mark = star , red!90]
    table[y=RsymL2G2K2t1cov,x=SNR]{Figs/data_Fig3.tex};
    \addlegendentry{\tiny Covariance-based-\textit{L}=2, \textit{G}=2, $\Omega$=2 }

    \addplot
    [dash dot, mark = o , red!90]
    table[y=RsymLMMSEmultistreamL2G2K2t1,x=SNR]{Figs/data_Fig3.tex};
    \addlegendentry{\tiny LMMSE-Multistream-\textit{L}=2, \textit{G}=2, $\Omega$=2, \textit{q}=2}

    \addplot
    [mark = square , red!90]
    table[y=RsymLMMSEsinglestreamL2G2K2t1,x=SNR]{Figs/data_Fig3.tex};
    \addlegendentry{\tiny LMMSE-Multistream-\textit{L}=2, \textit{G}=2, $\Omega$=2, \textit{q}=1}

    \addplot
    [dashed, mark = star , blue!90]
    table[y=RsymCovL5G4K3t1,x=SNR]{Figs/data_Fig3.tex};
    \addlegendentry{\tiny Covariance-based-\textit{L}=5, \textit{G}=4, $\Omega$=3}

\addplot
    [mark = square, blue!90]
    table[y=RsymproposedLMMSEsinglestreamL5G4K3t1,x=SNR]{Figs/data_Fig3.tex};
    \addlegendentry{\tiny LMMSE-Multistream-\textit{L}= 5, \textit{G}=4, $\Omega$=3, \textit{q}=1}

    \addplot
    [dash dot, mark = o , blue!90]
    table[y=RsymproposedLMMSEL5G4K3t1,x=SNR]{Figs/data_Fig3.tex};
    \addlegendentry{\tiny LMMSE-Multistream-\textit{L}= 5, \textit{G}=4, $\Omega$=3, \textit{q}=2}

     \addplot
    [dashed, mark = star , black!90]
    table[y=RsymCovL8G4K3t1,x=SNR]{Figs/data_Fig3.tex};
    \addlegendentry{\tiny Covariance-based-\textit{L}= 8, \textit{G}=4, $\Omega$=3 }

    \addplot
    [dash dot, mark = o , black!90]
    table[y=RsymLMMSEmultistreamL8G4K3t1,x=SNR]{Figs/data_Fig3.tex};
    \addlegendentry{\tiny LMMSE-Multistream-\textit{L}= 8, \textit{G}=4, $\Omega$=3, \textit{q}=2}

    \addplot
    [mark = square, black!70]
    table[y=RsymLMMSEsinglestreamL8G4K3t1,x=SNR]{Figs/data_Fig3.tex};
    \addlegendentry{\tiny LMMSE-Multistream-\textit{L}= 8, \textit{G}=4, $\Omega$=3, \textit{q}=1}




    
    
    \end{axis}

    \end{tikzpicture}
    }

    \caption{Proposed iterative solution for large DoF setup; $t=1,  K=10$}
    \label{fig:plot_3}
\end{figure}

In the scenarios depicted in the previous figures, we maintained $\beta = \binom{\Omega -1}{t}$, implying $q = 1$. Now, in the next simulation setup, we explore cases where $\beta$ surpasses $\binom{\Omega -1}{t}$, necessitating $q > 1$ to minimize the DoF loss.
In Figure~\ref{fig:plot_3}, we scrutinize various scenarios, encompassing both ($L = G =2$, $\Omega = 2$) and ($L = \{5, 8\}$, $G = 4$, $\Omega = 3$), all with a common CC gain of $t = 1$.
In the scenario where $L = G = 2$ and $\Omega = 2$, we observe a performance closely emulating the upper bound. Furthermore, in the case of ($L = 8$, $G = 4$, $\Omega = 3$), the proposed linear approach exhibits comparable performance to the covariance-based design, concurrently reducing computational overhead significantly. 
%
\par To gain a deeper understanding of the proposed schemes, in the same figure, we opt for an alternative comparison by contrasting our approach with constant $q = 1$. 
By opting for this as a baseline scheme~\cite{shariatpanahi2016multi}, we not only observe a decline in performance but also encounter a significant gap when compared to the maximum achievable DoF.
This behavior is more pronounced in the case of ($L = 8$, $G = 4$, $\Omega = 3$) compared to ($L = G =2$, $\Omega = 2$). This heightened intensity is attributed to the variation in the number of required parallel streams $\beta$, which decreases from $4$ to $2$ and results in a reduction in the total DoF from $6$ to $2$, respectively. 
On the contrary, in the configuration with ($L = 5$, $G = 4$, $\Omega = 3$), the performance is notably inferior compared to the upper bound, failing to match its DoF bound. Additionally, it closely aligns with a similar setup when $q$ is set to 1~\cite{shariatpanahi2016multi}.
This highlights a significant insight: when the transmitter-side spatial multiplexing ($L$) is insufficient w.r.t~\eqref{eq:total_DoF}, both the maximum attainable DoF and symmetric rate performance undergo degradation.
 In particular, since we assumed to achieve the maximum DoF of $4$ per user, we could not obtain this DoF via \eqref{eq:total_DoF} by $ L = 5$.  The maximum DoF is achievable in all other cases by the well-chosen transmitter spatial multiplexing gain, $L = 2$ for ($L = G = 2$, $\Omega = 2$) and $L = 8$ for ($L = 8$, $G = 4$ $\Omega = 3$).

\begin{figure}[ht]
    \centering
    \resizebox{0.84\columnwidth}{!}{%

    \begin{tikzpicture}

    \begin{axis}
    [
    axis lines = center,
    xlabel near ticks,
    xlabel = \smaller {SNR [dB]},
    ylabel = \smaller {Symmetric Rate [bits/s]},
    ylabel near ticks,
    ymin = 0,
    xmax = 30,
    legend pos = north west,
    ticklabel style={font=\smaller},
    grid=both,
    major grid style={line width=.2pt,draw=gray!30},
    ]
    
    
    
    

    




    

     \addplot
    [dashed, mark = o, red]
    table[y=RsymLG32K3t1ISITcov,x=SNR]{Figs/data.tex};
    \addlegendentry{\tiny Covariance-based-$L = 3$ }
    
    \addplot
    [mark = square, red]
    table[y=RsymLG32K3MMSESIC,x=SNR]{Figs/data.tex};
    \addlegendentry{\tiny  CVX-MMSE-SIC-$L = 3$}
    
    \addplot
    [dashed, mark = +, red]
    table[y=RsymLG32K3LMMSE,x=SNR]{Figs/data.tex};
    \addlegendentry{\tiny  CVX-LMMSE-$L = 3$}

     \addplot
    [only marks,mark = pentagon, red]
    table[y=RsymLG32K3LMMSEKKT,x=SNR]{Figs/data.tex};
    \addlegendentry{\tiny KKT-LMMSE-$L = 3$ }
    
    \addplot
    [dash dot, mark = pentagon, red]
    table[y=RsymLG32K3ZF,x=SNR]{Figs/data.tex};
    \addlegendentry{\tiny Zero-forcing-$L = 3$}

    \addplot
    [mark = star, black]
    table[y=RsymLG42K3t1ISITcov,x=SNR]{Figs/data.tex};
    \addlegendentry{\tiny Covariance-based-$L = 4$}

    \addplot
    [dashed, mark = diamond, black]
    table[y=RsymLG42K3MMSESIC,x=SNR]{Figs/data.tex};
    \addlegendentry{\tiny  CVX-MMSE-SIC-$L = 4$}

     \addplot
    [dashed, mark = +, black]
    table[y=RsymLG42K3LMMSE,x=SNR]{Figs/data.tex};
    \addlegendentry{\tiny CVX-LMMSE-$L = 4$}

    \addplot
    [only marks, mark = pentagon, black]
    table[y=RsymLG42K3LMMSEKKT,x=SNR]{Figs/data.tex};
    \addlegendentry{\tiny KKT-LMMSE-$L = 4$ }

    \addplot
    [dash dot, mark = halfcircle, black]
    table[y=RsymLG42K3ZF,x=SNR]{Figs/data.tex};
    \addlegendentry{\tiny Zero-forcing-$L = 4$}

    \end{axis}

    \end{tikzpicture}
    }


     \caption{Comparing performance of baseline and the proposed schemes; $t=1, G = 2, \Omega = 3, K=10$ }
    \label{fig:plot_1}
\end{figure}
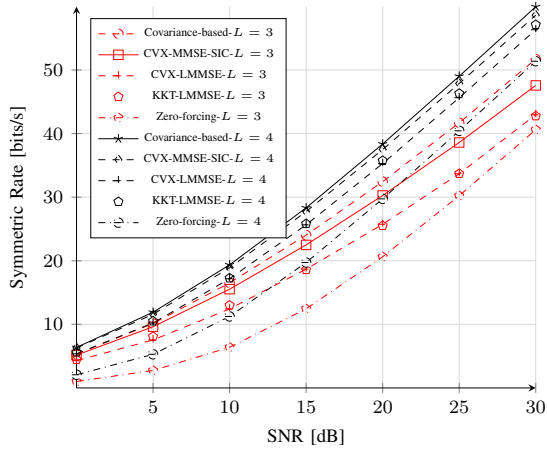

\begin{algorithm}[t] \label{F2}
	\caption{Iterative algorithm for \eqref{eq:optmization_problem} }
		\small\begin{enumerate}
		\item \textbf{Initialize} 
		$\lambda_{k,\CT}^i , t_{k,\CT}^i, {v}_{k,\CT}^i, \mu, \Bw_{\CT}^i, r_c , k \in \CK, \CT \in {\SfS}_k^{\CK}, i \in [q] $.
        set ${i}_1 =  0$, ${\lambda_{k,\CT}^i}^{(0)} = \frac{1}{K}$, ${v_{k,\CT}^i}^{(0)} = \frac{1}{K}$, $\Bw_{\CT}^i$ random vectors to satisfy power budget
		\item \textbf{Repeat} till convergence is met and ${i}_1 < I_{out}$, set ${i}_1 = {i}_1 + 1$, $i_1  = 0$; Compute ${\Bu_{k,\CT}^i}^{(0)}$ from~\eqref{eq:receive_beamformer} 
            \item \textbf{Repeat} till convergence is met and $i_2 < I_{in}$, set ${i}_2 = {i}_2 + 1$,
		\item Solve $\Bw_{\CT}^i$ from (\refeq{eq:primal_wg}), using ${\Bu_{k,\CT}^i}^{({i}_1-1)}$, ${\lambda_{k,\CT}^i}^{(i_2-1)}$, $\mu$. 
        \begin{itemize}
         \item[*] $\mu$ computed from bisection, meeting power budget
        \end{itemize}
        \item Compute $\epsilon_{k,\CT}^i$, using $\Bw_\CT^i$, ${\Bu_{k,\CT}^i}^{({i}_1-1)}$. 
        \item Compute $r_{c}$ from (\refeq{eq:primal_rc}), using $\epsilon_{k,\CT}^i$, ${v_{k,\CT}^i}^{(i_2-1)}$.
        \item Update ${v_{k,\CT}^i}^{(i_2)}$ from (\refeq{eq:subgradient_vkl}), using $r_{c}$,  ${v_{k,\CT}^i}^{(i_2-1)}$.
        \item Normalize ${v_{k,\CT}^i}^{(i_2)}$ by $\frac{1}{q}\sum_{\CT \in {\SfS}_k^{\CK}, i \in [q] } {v}_{k,\CT}^{i}$; to satisfy (\refeq{eq:dual_vkl_sum})
		\item Update ${\lambda_{k,\CT}^i}^{(i_2)}$\!\! from \eqref{eq:dual_lambda}, using ${{v}_{k,\CT}^i}^{(i_2)}$,\!\! $\epsilon_{k,\CT}^i$ 10) \textbf{end}, 11) \textbf{end} 
	\end{enumerate}
\end{algorithm}

\section{Conclusion and Future Work}

In our study of multicast beamformer design for MIMO-coded caching systems, we introduced a versatile approach enabling flexible multicast transmissions to partially overlapping groups of users. Overcoming practical limitations, our design ensures enhanced DoF and system performance, compared to state-of-the-art, across various MIMO-CC configurations. We assessed its effectiveness by formulating the symmetric rate as the sum rate of sub-streams received across partially overlapping streams per user, utilizing a linear multi-stream multicast transmission vector and an LMMSE receiver. Finally, to overcome the non-convex nature of the resulting problem, we used alternative optimization and SCA techniques, supported by a rapid iterative Lagrangian-based algorithm. 


\bibliographystyle{IEEEtran}
\bibliography{references,whitepaper}


\end{document}